# Ethical AI in Retail: Consumer Privacy and Fairness
Anthonette Adanyin
A.adanyin@wlv.ac.uk
15 blackthorn drive wolverhampton wv10 8an united kingdom

**Abstract**

The adoption of artificial intelligence (AI) in retail has significantly transformed the industry, enabling more personalized services and efficient operations. However, the rapid implementation of AI technologies raises ethical concerns, particularly regarding consumer privacy and fairness. This study aims to analyze the ethical challenges of AI applications in retail, explore ways retailers can implement AI technologies ethically while remaining competitive, and provide recommendations on ethical AI practices. A descriptive survey design was used to collect data from 300 respondents across major e-commerce platforms. Data were analyzed using descriptive statistics, including percentages and mean scores. Findings shows a high level of concerns among consumers regarding the amount of personal data collected by AI-driven retail applications, with many expressing a lack of trust in how their data is managed. Also, fairness is another major issue, as a majority believe AI systems do not treat consumers equally, raising concerns about algorithmic bias. It was also found that AI can enhance business competitiveness and efficiency without compromising ethical principles, such as data privacy and fairness. Data privacy and transparency were highlighted as critical areas where retailers need to focus their efforts, indicating a strong demand for stricter data protection protocols and ongoing scrutiny of AI systems. The study concludes that retailers must prioritize transparency, fairness, and data protection when deploying AI systems. The study recommends ensuring transparency in AI processes, conducting regular audits to address biases, incorporating consumer feedback in AI development, and emphasizing consumer data privacy.

## 1.0 Introduction

The adoption of artificial intelligence (AI) in retail has fundamentally transformed the industry, enabling retailers to offer more personalized and efficient services while optimizing supply chains and inventory management. AI-powered systems now play a critical role in predicting consumer behavior, customizing product recommendations, and automating customer service. In a study by Shekhawat (2022), 75% of retailers have reported adopting some form of AI technology by 2021, with AI investments expected to reach $7.3 billion annually by 2022. This trend is driven by the growing demand for seamless, personalized shopping experiences, with 85% of customer interactions expected to be managed without human agents by 2025 (Guha et al., 2021). AI's ability to process and analyze vast amounts of consumer data has

revolutionised personalisation in retail. Algorithms used by companies like Amazon and Alibaba analyze consumer data to provide tailored recommendations and offers. Study has shown that Alibaba's AI-powered voice assistant led to a 23% increase in consumer spending and an 11% rise in browsing, contributing to an additional $613 million in annual sales (Sun et al., 2019). Moreover, 45% of global consumers reported that they were more likely to engage with retailers offering personalized experiences, further highlighting the importance of AI in maintaining competitiveness (Guha et al., 2021).

However, this rapid adoption of AI comes with significant ethical concerns, particularly regarding consumer privacy. According to Wang et al. (2021), 90% of consumers expressing concerns about how their personal data is collected and used, data privacy has become a major issue in AI adoption. AI systems rely on vast datasets that include sensitive information, such as browsing history, purchase behavior, and even biometric data. This raises concerns about potential misuse or breaches. In a study, 78% of consumers reporting a lack of trust in how companies handle their data (Guha et al., 2021).

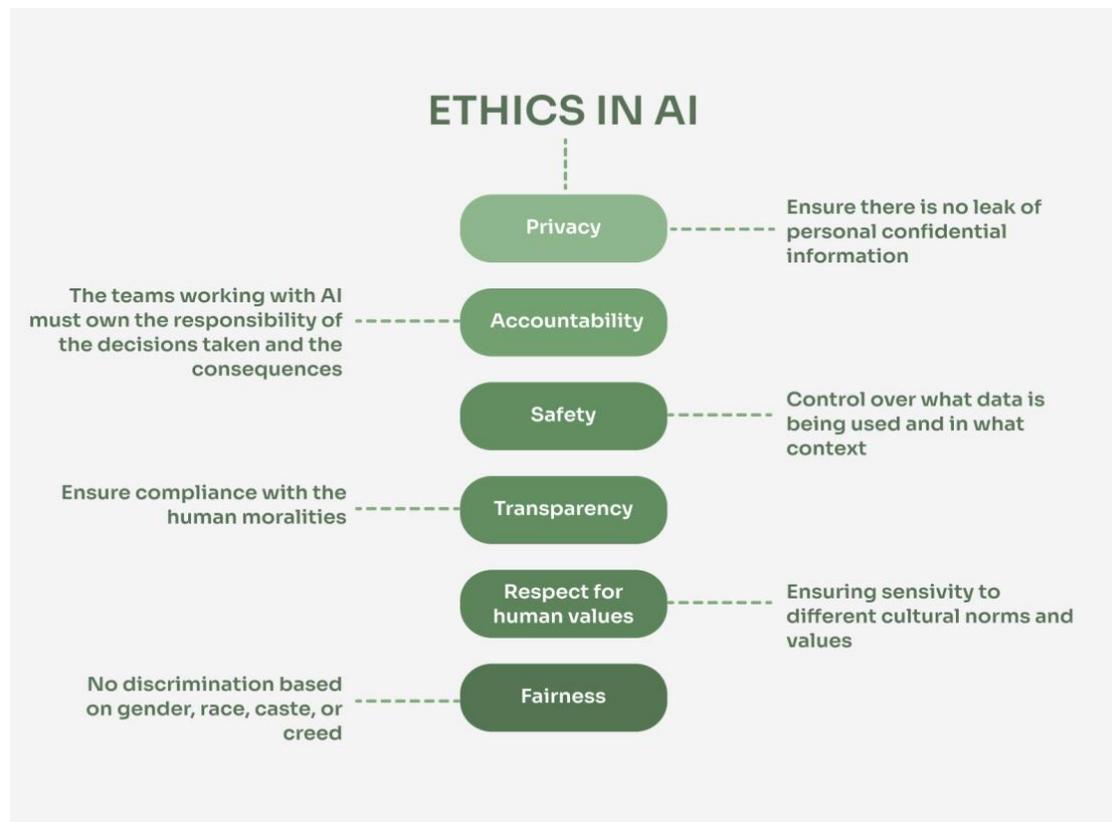

**Figure 1: Ethic considerations in AI (Sharma et al., 2024)**

Furthermore, fairness in AI decision-making has become another critical concern. Algorithms used in customer segmentation and product recommendations are often opaque, leading to potential biases that disproportionately impact certain demographic groups. Studies have shown that 56% of AI systems used in retail may reinforce pre-existing biases, leading to discriminatory outcomes in pricing, product offerings, and marketing (Pizzi & Scarpi, 2020). These biases not only harm consumer trust but also expose retailers to legal and reputational risks. In response to these ethical challenges, regulators and retailers are increasingly focusing on transparency and fairness in AI usage. Report shows that 44% of retailer surveyed in 2021 stated that they are actively working to improve algorithmic transparency and reduce biases in their AI systems (Tiutiu & Dabija, 2023). Companies are also investing in secure data management systems to ensure compliance with privacy regulations, such as the General Data Protection Regulation (GDPR) in Europe, which mandates strict guidelines on consumer consent and data usage.

## 1.1 Statement of Problem

While AI-driven retail offers significant benefits, it also introduces risks to consumer privacy and fairness. Ethical dilemmas include biased algorithms, the potential misuse of personal data, and a lack of transparency in decision-making processes. Retailers face the challenge of implementing AI systems that respect consumer rights while maintaining operational efficiency. The absence of comprehensive ethical frameworks for AI in retail creates a pressing need to address these concerns (Martin et al., 2020). Hence this study.

## 1.2 Objectives of the Study

The specific objectives of this study are to:

i. Analyze the ethical challenges surrounding AI applications in retail, particularly in relation to consumer privacy and fairness.

ii. Explore how retailers can implement AI technologies that align with ethical principles while maintaining business competitiveness.

iii. Provide practical recommendations for retailers on developing ethical AI systems.

## 2.0 Literature Review

Ethical AI refers to the development and deployment of artificial intelligence systems that respect human values, societal norms, and fundamental rights (Kaplan, 2020).

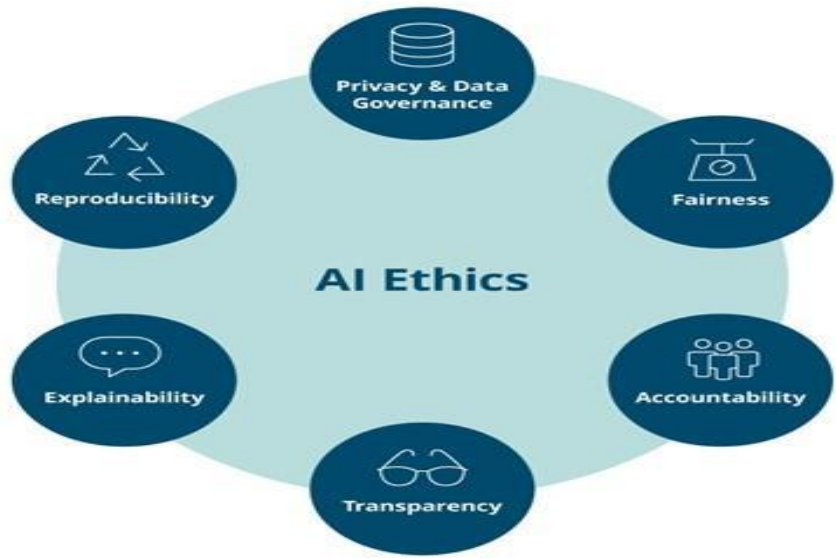

**Figure 2: AI Ethics (Sharma et al., 2024)**

As illustrated in the image, key principles of ethical AI include Privacy & Data Governance, Fairness, Accountability, Transparency, Explainability, and Reproducibility. These principles are especially important in retail, where maintaining consumer trust is critical to long-term success.

**Privacy and Data Governance** is a fundamental aspect of ethical AI. In retail, this means ensuring that consumer data is collected, stored, and processed securely and in compliance with privacy regulations. Retailers must ensure that customers give informed consent for their data to be used and that there are robust protections in place to prevent unauthorized access or data misuse. This is vital because any breach of consumer data can result in a loss of trust and serious legal implications (Aldboush & Ferdous, 2023). Managing this data responsibly is essential to maintaining consumer trust. Privacy concerns arise when there is a lack of transparency about how data is collected, stored, and used, or when data is shared without consumer consent. Retailers must adopt ethical guidelines that emphasize data protection, informed consent, and transparency in data usage (Pizzi & Scarpi, 2020)

**Fairness** in AI ensures that all consumers are treated equally, without discrimination based on race, gender, age, or other characteristics. In the retail context, fairness implies that AI-driven decisions, such as personalized recommendations, pricing, and customer profiling, are free from discrimination and do not reinforce existing social inequalities. Retail AI systems, such as those used for personalized marketing or pricing, must be free from bias. However, AI systems trained on biased data can perpetuate unfair treatment, making it imperative for retailers to address algorithmic bias and ensure that AI decision-making processes are transparent and equitable. AI systems should undergo continuous monitoring to ensure that they do not unfairly disadvantage certain groups of consumers (Chadha, 2024). Ensuring fairness is not only an ethical requirement but also helps in maintaining a diverse customer base and avoiding potential legal challenges. Studies have shown that fairness concerns are especially prominent when AI decisions affect personalized marketing and customer engagement strategies (Cheung & To, 2020).

**Accountability** refers to the responsibility of businesses to ensure that their AI systems operate ethically. In the retail context, companies must take responsibility for the actions and outcomes produced by their AI systems, ensuring that any negative consequences, such as bias or data misuse, are promptly addressed. Retailers should implement clear procedures to monitor and rectify AI-related issues to preserve consumer trust (Sheriffdeen, 2024).

**Transparency** is crucial for fostering trust in AI-driven retail operations. Consumers need to understand how their data is being used, how AI makes decisions, and what impact it has on their shopping experience. Retailers should provide clear and accessible information about how AI systems function, particularly when it comes to data collection and personalized recommendations. Transparency can help alleviate consumer concerns about AI's role in their shopping experiences and foster greater trust (Sheriffdeen, 2024).

**Explainability** is closely related to transparency. AI systems should be designed so that their decisions can be easily explained to consumers (Kim et al., 2023). When a retail AI system recommends a product or adjusts prices based on customer behavior, consumers should be able to understand why those decisions were made. This can be done through clear communication about the AI's decision-making processes, helping to build consumer confidence in the fairness and accuracy of AI systems (de Fine Licht & de Fine Licht, 2020).

**Reproducibility** ensures that AI systems deliver consistent and reliable results across different scenarios (Li et al., 2023). Retailers need to ensure that the outcomes generated by their AI systems are not random or overly influenced by one-off events. By ensuring that AI systems perform predictably and reliably, retailers can maintain consistency in customer experiences and prevent unexpected outcomes that could harm consumer trust.

The intersection of ethical AI, fairness, and privacy is crucial for responsible AI deployment in retail. Ethical AI must ensure that fairness and privacy are upheld throughout AI applications. This requires continuous monitoring of AI systems to eliminate biases and safeguard consumer data, while adhering to ethical guidelines that promote transparency and accountability (Sharma et al., 2023)

**2.2 Emprircal Review**

Pillai et al. (2020) investigated the factors influencing consumers' intentions to shop at AI-powered automated retail stores (AIPARS). The study utilized the Technology Readiness and Acceptance Model (TRAM) to survey 1,250 consumers, analyzing the data with Partial Least Squares Structural Equation Modeling (PLS-SEM). The research revealed that perceived usefulness, fairness, perceived enjoyment, and customization were significant predictors of consumers' shopping intentions at AI-powered stores. Innovativeness and optimism positively affected consumer attitudes, while insecurity had a negative impact on perceived usefulness. Tiutiu & Dabija (2023) examined the impact of artificial intelligence on customer experience enhancement in online retail, focusing on ethical aspects. The study was based on quantitative research through an online survey of 272 participants. The data was analyzed using regression analysis to test the conceptual model. The study found that AI can significantly improve customer experiences by providing safe, customer-friendly, and ethically applied technology. Ethical AI implementation was identified as a key factor for building consumer trust and loyalty. Stanciu & Rîndașu (2021) explored privacy concerns related to AI-driven retail mobile applications, particularly focusing on how excessive data collection affects consumer trust. The researchers examined 117 mobile shopping applications available in the Google Play store, analyzing the permissions and personal data collected by each app. The study found that many AI-driven mobile applications intrude on consumer privacy by collecting excessive personal information. It emphasized the need for stricter privacy regulations and transparent AI usage to maintain consumer trust.

Pizzi & Scarpi (2020) investigated how new retail technologies, including AI, influence consumer privacy perceptions and trust in retailers. The authors conducted a moderated serial mediation analysis, focusing on distributive fairness and the technology's hedonism as antecedents of consumer privacy perceptions. They found that privacy perceptions were significantly affected by distributive fairness, and that perceived hedonism positively influenced technology acceptance. Trust in retailers was also found to be critical for managing privacy concerns. Guha et al. (2021) examined the broader implications of AI in retail, focusing on its impact on value creation and ethical concerns. The study involved interviews with senior retail managers to identify how AI applications, both customer-facing and non-customer-facing, influence retail operations. The findings suggested that AI can improve retail efficiency, but there are significant ethical concerns regarding privacy and transparency. AI was found to be most effective when augmenting rather than replacing human decision-making.

Sharma et al. (2023) examined the ethical dilemmas involved in AI-based marketing, particularly how businesses balance profitability with consumer trust. The authors conducted a comprehensive literature review and case studies of AI-driven marketing practices, focusing on data privacy, security, and the ethical use of consumer data. The study highlighted that transparency in AI-based marketing practices and responsible data usage were essential for maintaining consumer trust. Striking a balance between hyper-targeted marketing and respecting consumer privacy emerged as a critical challenge. Martin et al. (2020) explored the complex interplay between consumer privacy concerns, regulatory influences, and retail strategies. The study utilized a multi-method approach, combining expert interviews, a large-scale consumer survey, and global case studies to examine privacy concerns in the retail industry. The research identified that big data is a key driver of customer relationship performance but also highlighted the profound impact of regulations on retail data practices. The study emphasized that proactive privacy protection strategies were essential for maintaining consumer trust.

**Contribution of This Study**

Although several studies highlight ethical issues in AI, there is limited research specifically focusing on how retailers can balance the ethical trade-offs between privacy, fairness, and profitability. This study fills that gap by offering a comprehensive framework for retailers to address these ethical concerns while maintaining competitive advantage.

**3.0 Methodology**

This study employs a descriptive survey research design to investigate ethical AI in retail. A descriptive survey research design was chosen to due to its ability to capture detailed, quantitative data from a broad audience efficiently (Siedlecki, 2020). For the purpose of this study, the population was made up of three hundred (300) purposefully targeted respondents across major e-commerce platforms. The study adopts a purposive sampling technique which allows the researcher to select individuals who are most likely to provide relevant and rich data specific to the study's objectives. Online surveys (google forms) was distributed to targeted respondents. The survey was a structured close-ended called Ethical AI in Retail: Consumer Privacy and Fairness (EAR-CPF).

Respondents were provided with ethical consideration forms to ensure their permission and consent were obtained prior to sending the online surveys. Each targeted respondent received an email containing the survey links. The questions were analysed by utilizing descriptive statistics, including percentages, means, and standard deviations.

**4.0 Results**

**Table 1: Ethical Challenges Surrounding AI Applications In Retail**

| Items | SA (%) | A (%) | D (%) | SD (%) | M | Std. D | Remarks |
|---|---|---|---|---|---|---|---|
| **Privacy** | | | | | | | |
| 1. I am concerned about how much personal data AI-driven retail applications collect from me. | 43 (53.2) | 36 (45.6) | 1 (1.1) | 0 (0) | 3.52 | 0.52 | Strongly Agree |
| 2. AI systems used by retailers are transparent about how they use my personal data. | 0 (0) | 0 (0) | 167 (55.6) | 133 (44.4) | 1.56 | 0.49 | Disagree |
| 3. Retailers using AI in their services take appropriate measures to protect my personal data from misuse or breaches | 34 (42.2) | 23 (28.9) | 18 (22.2) | 5 (6.1) | 3.07 | 0.95 | Agree |
| **Fairness** | | | | | | | |
| 4. AI-driven retail systems treat all consumers equally, without favoring certain groups or demographics | 0 (0) | 0 (0) | 127 (42.2) | 162 (53.9) | 1.50 | 0.57 | Disagree |
| 5. I believe that AI | 0 | 36 | 57 | 183 | 1.60 | 0.80 | Disagree |
| systems used in retail make unbiased decisions when recommending products or services. | (0) | (20) | (18.9) | (61.1) | | | |
| 6. Retailers should provide clear explanations of how AI systems make decisions that affect my shopping experience | 177 (58.9) | 115 (38.3) | 8 (2.8) | 0 (0) | 3.56 | 0.55 | Strongly Agree |

*Strongly Agree (SA) =4, Agree (A) = 3, Disagree (D) = 2, Strongly Disagree (SD) = 1, St.D = Standard Deviation, M= Mean*

Table 1 shows responses on ethical challenges surrounding AI applications in retail, focusing on consumer privacy and fairness concerns. For privacy, 53.2% of respondents strongly agreed, and 45.6% agreed that they are concerned about the amount of personal data collected by AI-driven retail applications, reflecting a high level of concern (M = 3.52, Std. D = 0.52). However, when asked if AI systems used by retailers are transparent about how they use personal data, none of the respondents agreed or strongly agreed. Instead, 55.6% disagreed and 44.4% strongly disagreed, indicating a significant lack of transparency (M = 1.56, Std. D = 0.49). Regarding data protection, 42.2% strongly agreed and 28.9% agreed that retailers take appropriate measures to safeguard data, while 22.2% disagreed, and 6.1% strongly disagreed (M = 3.07, Std. D = 0.95), showing moderate confidence in retailers' data protection practices.

For fairness, none of the respondents felt that AI-driven retail systems treat consumers equally. A majority, 53.9%, strongly disagreed, and 42.2% disagreed, indicating significant concerns about biased treatment in AI systems (M = 1.50, Std. D = 0.57). Similarly, 61.1% strongly disagreed, and 18.9% disagreed that AI systems make unbiased decisions, suggesting prevalent concerns about bias in AI-driven recommendations (M = 1.60, Std. D = 0.80). Lastly, 58.9% strongly agreed, and 38.3% agreed that retailers should provide clear explanations of how AI decisions affect their shopping experience, emphasizing the demand for greater transparency (M = 3.56, Std. D = 0.55). Overall, the analysis shows substantial concerns about privacy, transparency, and fairness in AI applications in retail, with a strong demand for transparency and fair treatment from consumers.

**Table 2: Implementation of AI technologies that align with ethical principles while maintaining business competitiveness**

|   | Items | SA (%) | A (%) | D (%) | SD (%) | M | Std. D | Remarks |
|---|---|---|---|---|---|---|---|---|
| 1 | Retailers can balance profitability and ethical principles by using AI technologies in a transparent manner, ensuring consumer trust | 93 (31.1) | 172 (57.2) | 35 (11.7) | 0 (0) | 3.19 | 0.63 | Agree |
| 2. | The integration of AI in retail can enhance business efficiency while adhering to ethical standards, such as protecting consumer data privacy | 152 (50.6) | 137 (45.6) | 12 (3.9) | 0 (0) | 3.47 | 0.57 | Agree |
| 3. | AI technologies help retailers achieve competitiveness while ensuring fairness in customer interactions, such as preventing algorithmic bias. | 95 (31.7) | 205 (68.3) | 0 (0) | 0 (0) | 3.31 | 0.47 | Agree |
| 4. | Retailers that implement AI technologies ethically (e.g., by providing transparent AI-driven decisions) are more likely to maintain customer loyalty and business success. | 185 (61.7) | 107 (35.6) | 3 (1.1) | 5 (1.7) | 3.57 | 0.61 | Strongly Agree |
| 5 | AI-driven automation in retail can improve | 24 (30) | 56 (70) | 0 (0) | 0 (0) | 3.30 | 0.46 | Agree |

| | operational efficiency without compromising ethical standards, such as fair treatment of employees and customers | | | | |
|---|---|---|---|---|---|

Table 2 shows responses on the implementation of AI technologies that align with ethical principles while maintaining business competitiveness. For the first item, 31.1% of respondents strongly agree and 57.2% agree that retailers can balance profitability and ethical principles through transparent AI technologies, ensuring consumer trust. With a mean score of 3.19, indicating general agreement, though 11.7% express concerns with transparency.

Regarding the integration of AI in retail to enhance business efficiency while adhering to ethical standards, such as protecting consumer data privacy, majority strongly agree (50.6%) or agree (45.6%), resulting in a high mean score of 3.47. Only 3.9% disagree, suggesting widespread confidence in AI's ability to balance efficiency and ethics.

On competitiveness and fairness, 31.7% strongly agree and 68.3% agree that AI technologies help retailers achieve competitiveness while ensuring fairness in customer interactions, such as preventing algorithmic bias. This leads to a mean score of 3.31, reflecting overall agreement. In terms of customer loyalty and business success, 61.7% strongly agree and 35.6% agree that ethically implemented AI (e.g., providing transparent AI-driven decisions) maintains customer loyalty, giving a mean score of 3.57, demonstrating strong support for ethical AI practices. Finally, on the question of AI-driven automation improving operational efficiency without compromising ethical standards, 30% strongly agree and 70% agree, with no disagreement, yielding a mean score of 3.30, indicating broad support for automation that upholds ethical treatment. Overall, the responses show strong agreement that AI can enhance competitiveness, efficiency, and fairness while maintaining ethical principles in retail, particularly through transparency and fair treatment.

**Table 3: Practical recommendations for retailers on developing ethical AI systems.**

|   | Items | SA (%) | A (%) | D (%) | SD (%) | M | Std. D | Remarks |
|---|---|---|---|---|---|---|---|---|
| 1 | Retailers should ensure that their AI systems are transparent and provide clear information on how decisions are made | 26 (32.2) | 55 (44) | 8 (9.4) | 3 (3.3) | 3.16 | 0.73 | Agree |
| 2. | AI systems in retail should be regularly audited to ensure fairness and prevent algorithmic biases | 21 (26.1) | 55 (68.3) | 4 (5.6) | 0 (0) | 3.21 | 0.52 | Agree |
| 3. | Retailers should incorporate consumer feedback into the development of their AI systems to ensure they meet ethical standards | 16 (20) | 62 (77.8) | 2 (2.2) | 0 (0) | 3.18 | 0.44 | Agree |
| 4. | Ethical AI systems should prioritize consumer data privacy and security in all retail applications. | 52 (65) | 23 (28.3) | 4 (4.4) | 2 (2.2) | 3.65 | 0.69 | Strongly Agree |
| 5 | AI-driven automation in retail should be designed to support fairness and equal treatment for all consumers. | 27 (33.9) | 48 (59.4) | 5 (6.7) | 0 (0) | 3.27 | 0.58 | Agree |

Table 3 presents responses to practical recommendations for retailers on developing ethical AI systems. From the table 32.2% of respondents strongly agree and 44% agree that retailers should ensure their AI systems provide clear information on how decisions are made, leading to a mean score of 3.16. However, 9.4% disagree and 3.3% strongly disagree, indicating some reservations about the current level of transparency in AI systems. Regarding regular audits to ensure fairness and prevent algorithmic biases, 26.1% strongly agree and 68.3% agree, resulting in a mean score of 3.21, showing strong support for regular audits. Only 5.6% disagree, indicating minimal opposition to this practice. For incorporating consumer feedback into AI development, 20% strongly agree and 77.8% agree, with only 2.2% disagreeing. This results in a mean score of 3.18, indicating broad agreement on the importance of consumer feedback to meet ethical standards. On prioritizing consumer data privacy and security, 65% strongly agree and 28.3% agree, reflecting strong agreement with a mean score of 3.65 on the importance of data privacy in ethical AI systems. Lastly, for AI-driven automation supporting fairness and equal treatment, 33.9% strongly agree and 59.4% agree, with only 6.7% disagreeing, resulting in a mean score of 3.27, demonstrating overall agreement on the need for fairness in AI-driven automation.

**5.0 Discussions**

The findings from the analysis provide significant insights into the ethical challenges and implementation strategies for AI in retail, particularly concerning consumer privacy, fairness. The high levels of concern about the amount of personal data collected by AI-driven retail applications suggest that consumers are wary of how their data is being managed. This is consistent with a study Pizzi & Scarpi (2020), which highlighted consumer apprehension about privacy in AI applications in retail. However, the lack of transparency in how retailers use personal data is alarming. This finding is inline with a research by Wang et al. (2021), who emphasized the importance of transparency in AI to maintain consumer trust. In terms of data protection, while a moderate level of confidence is indicated that retailers take appropriate measures to safeguard data, reflecting a cautious optimism similar to what Guha et al. (2021) discussed in their study on AI's role in enhancing retail efficiency without compromising ethical standards. Findings also shows an overwhelming perception that AI systems do not treat consumers fairly aligns with concerns about algorithmic bias in retail. None of the respondents felt that AI systems treated consumers equally indicating a significant gap in fairness. This concern is corroborated by Cheung & To (2020), which emphasized the potential for AI to perpetuate biases in customer interactions and pricing. Also the findings that retailers need to provide clearer explanations of how AI decisions affect shopping experiences aligns with Pillai et al. (2020), who found that transparency in AI-driven decisions improves customer trust and satisfaction.

Further, findings suggest that consumers believe AI can enhance both competitiveness and ethical standards, particularly when it comes to balancing profitability with transparency. This supports the argument by Cao (2021) that transparency in AI applications is crucial for maintaining both competitive advantage and ethical standards. Additionally, a strong majority believe that AI can improve business efficiency while adhering to ethical principles like data privacy. This is corroborated the findings by Anica-Popa et al. (2021), who showed that AI-driven efficiency can be achieved without sacrificing ethical considerations like privacy. In terms of competitiveness and fairness, there is broad agreement that AI technologies can help retailers remain competitive while preventing algorithmic bias. This is consistent with the findings of Stanciu & Rîndașu (2021), who explored how AI can enhance retail efficiency without compromising fairness. The high level of agreement that ethical AI maintains customer loyalty and business success further supports the research of Oosthuizen et al. (2020), who found that transparency and fairness in AI applications directly influence customer loyalty.

Additionally, the findings also indicates a strong support for practical measures to enhance the ethical use of AI in retail. A significant number of respondents agree that AI systems need to be transparent and provide clear information on decision-making processes. Kaplan (2020) that stresses the importance of transparency as an ethical guideline for AI development, particularly in consumer-facing applications . Furthermore, the recommendation for regular audits to ensure fairness and prevent bias is widely supported with minimal opposition, indicating a general agreement that ongoing scrutiny of AI systems is essential to maintaining fairness. Sharma et al. (2023) in their study advocated for continuous monitoring and auditing of AI systems to avoid ethical pitfalls like bias and discrimination. The emphasis on incorporating consumer feedback into AI development also aligns with a work by Floridi et al. (2018) who emphasized the role of consumer input in shaping AI systems that are both user-friendly and ethically aligned. The high level of agreement on data privacy, further reinforces the ongoing demand for strict data protection protocols in AI systems, which is consistent with the findings of Stanciu & Rîndașu (2021), who highlighted privacy as a primary concern for consumers in AI-driven retail applications.

**6.0 Conclusion**

The findings in this study underscore the need for retailers to prioritize transparency, fairness, and data protection when implementing AI systems. Consumer trust hinges on the ability of retailers to address privacy concerns and ensure that AI systems operate transparently and equitably. The support for regular audits and incorporating consumer feedback into AI development reflects a growing expectation that retailers will actively work to prevent bias and maintain ethical standards. The consistent demand for transparency, as highlighted across multiple items, suggests that retailers who fail to address these concerns may face consumer pushback and potential reputational damage. Moreover, the broad support for ethical AI practices, such as fair treatment and data security, indicates that aligning AI systems with ethical principles is not only a moral imperative but also a strategic one for maintaining customer loyalty and long-term business success. Transparency, fairness, and consumer data protection remain critical areas where retailers must focus their efforts to ensure that AI technologies are both competitive and ethically sound.

## 7.0 Recommendations

Based on the analysis, several recommendations were made. First, retailers should prioritize transparency in their AI systems by providing consumers with clear, accessible information about how data is collected and decisions are made. This transparency will not only foster consumer trust but also ensure that ethical standards are upheld in all AI-driven processes. Second, it is essential that retailers conduct regular audits of their AI systems to ensure fairness and prevent algorithmic biases. By continuously monitoring these systems, retailers can identify and address any unintended biases that could lead to unfair treatment of certain consumer groups, thereby safeguarding both ethical principles and business integrity. Third, retailers must actively seek and incorporate consumer feedback into the development and refinement of AI technologies. Engaging consumers in the development process will help ensure that AI systems meet ethical standards and align with consumer expectations, fostering a more inclusive and consumer-centric approach. Additionally, protecting consumer data privacy should remain a top priority in all AI applications. Retailers must implement stringent data protection measures and ensure compliance with regulations to maintain consumer confidence and mitigate the risks associated with data breaches or misuse. Finally, AI-driven automation in retail should be designed to promote fairness and equal treatment for all consumers, ensuring that automated processes do not inadvertently favor certain demographics over others. By committing to these ethical practices, retailers can effectively leverage AI technologies to enhance competitiveness while maintaining the trust and loyalty of their customers.